\documentclass[aps,pre,showpacs,amsmath,amsfonts,amssymb,notitlepage]{revtex4-1} 

\usepackage[dvips]{graphicx}
\usepackage{bbm}
\usepackage{xcolor}
\usepackage[colorlinks,linkcolor=blue,anchorcolor=blue,citecolor=blue,filecolor=blue,menucolor=blue,urlcolor=blue]{hyperref}
\usepackage{breakurl}

\newcommand{\beq}{\begin{equation}} 
\newcommand{\eeq}{\end{equation}}
\newcommand{\bea}{\begin{eqnarray}} 
\newcommand{\eea}{\end{eqnarray}}

\input epsf 

\begin{document} 

\title{Density of states of the $XY$ model: an energy landscape approach}

\author{Cesare Nardini}
\email{cesare.nardini@gmail.com}
\affiliation{Laboratoire de Physique, \'Ecole Normale Sup\'erieure de Lyon and CNRS, 46 all\'ee d'Italie, F-69007 Lyon, France}
\author{Rachele Nerattini}
\email{rachele.nerattini@fi.infn.it}
\affiliation{Dipartimento di Fisica e Astronomia and Centro per lo Studio
delle Dinamiche Complesse (CSDC), Universit\`a di
Firenze, and Istituto Nazionale di Fisica Nucleare (INFN), Sezione di
Firenze, via G.\ Sansone 1, I-50019 Sesto Fiorentino (FI), Italy}
\author{Lapo Casetti}
\email{lapo.casetti@unifi.it}
\affiliation{Dipartimento di Fisica e Astronomia and Centro per lo Studio
delle Dinamiche Complesse (CSDC), Universit\`a di
Firenze, and Istituto Nazionale di Fisica Nucleare (INFN), Sezione di
Firenze, via G.\ Sansone 1, I-50019 Sesto Fiorentino (FI), Italy}
\date{December 17, 2013}

\begin{abstract}
Among the stationary configurations of the Hamiltonian of a classical O$(n)$ lattice spin model, a class can be identified which is in one-to-one correspondence with all the the configurations of an Ising model defined on the same lattice and with the same interactions. Starting from this observation it has been recently proposed that the microcanonical density of states of an O$(n)$ model could be written in terms of the density of states of the corresponding Ising model. Later, it has been shown that a relation of this kind holds exactly for two solvable models, the mean-field and the one-dimensional $XY$ model, respectively. We apply the same strategy to derive explicit, albeit approximate, expressions for the density of states of the two-dimensional $XY$ model with nearest-neighbor interactions on a square lattice. The caloric curve and the specific heat as a function of the energy density are calculated and compared against simulation data, yielding a very good agreement over the entire energy density range. The concepts and methods involved in the approximations presented here are valid in principle for any O$(n)$ model. 
\end{abstract}

\keywords{Lattice spin models, density of states, phase transitions, energy landscapes}
\maketitle

\section{Introduction}\label{sec:Intro}
A quite recent point of view on the study of the thermodynamic properties of a classical $N-$body Hamiltonian system, commonly referred to as the ``energy landscape'' approach \cite{Wales:book}, implies the study of the properties of the graph of the energy 
function $H:\Gamma_{N}\rightarrow\mathbb{R}$, where $\Gamma_{N}$ is phase space of the system. Examples of applications include disordered systems and glasses 
\cite{DebenedettiStillinger:nature2001,Sciortino:jstat2005}, clusters \cite{Wales:book}, biomolecules and protein folding \cite{OnuchicLutheyWolynes:arpc1997,prl2006_2,pre2008}. In this context a special r\^{o}le is played by the stationary points of the energy function. The latter are the points $p_{c}$ in phase space such that $\nabla H(p_{c})=0$. 
Based on knowledge about the stationary points of the energy function, landscape methods allow to estimate both dynamical and equilibrium properties of a system. In the present paper the attention will be focused only on equilibrium properties. 

The natural setting to understand the connection between the stationary points of the Hamiltonian and equilibrium statistical properties is the microcanonical one \cite{Kastner:physa2006}. This is the statistical description in which the system is considered as isolated; the control parameter is the energy density $\varepsilon=E/N$, and the relevant thermodynamic potential is the entropy density $s_N$ given by  
\begin{equation}
 s_{N}(\varepsilon)=\frac{1}{N}\log \omega_{N}(\varepsilon)\,,
\end{equation}
where $\omega_{N}(\varepsilon)$ is the density of states of the system. The density of states is given by 
\begin{equation}\label{intro:coarea}
 \omega_{N}(\varepsilon)=\int_{\Gamma_N}d\Gamma_{N}\delta\left(H-N\varepsilon\right)=
 \int_{\Gamma_N\cap\Sigma_{\varepsilon}}\frac{d\Sigma_{\varepsilon}}{|\nabla H|},
 \end{equation}
where $\Sigma_{\varepsilon}$ is the hypersurface of constant energy $N\varepsilon$ and $d\Sigma$ is 
the Hausdorff measure; the rightmost integral stems from a co-area formula \cite{Federer:book}. 
When a stationary configuration $p_{c}$ is considered, the integrand in (\ref{intro:coarea}) 
diverges. In finite-$N$ systems this divergence is compensated by the measure $d\Sigma$ that shrinks in such a way that the 
integral in Eq.\ (\ref{intro:coarea}) remains finite. This notwithstanding, the density of states is non-analytic
at the corresponding value of the energy density $\varepsilon_{c}=H(p_{c})/N$. Although these nonanalyticities typically disappear in the thermodynamic limit $N\to\infty$, in some special cases they may survive and give rise to 
equilibrium phase transitions \cite{prl2006,KSS:jstat2008,jstat2009}. Stationary points of the energy correspond to topology changes in the accessible phase space, and this has suggested a relation between these topology changes and equilibrium phase transitions (see Refs.\ \cite{physrep2000,Pettini:book,Kastner:rmp2008} and references therein).

In \cite{prl2011} an approximate form for the density of states of a paradigmatic class of classical spin models, the O$(n)$ models, was conjectured on the basis of energy landscape considerations. According to the conjecture, the density of states $\omega^{(n)}$ of a classical O$(n)$ spin model on a lattice can be approximated in terms of the density of states $\omega^{(1)}$ of the corresponding Ising model, i.e., an Ising model defined on the same lattice and with the same interactions. The crucial observation is that all the configurations of the corresponding Ising model are stationary configurations of a O$(n)$ model for any $n$. The density of states would then be given by
\begin{equation}\label{intro:ansatz}
\omega^{(n)}(\varepsilon) \approx \omega^{(1)}(\varepsilon)g^{(n)}(\varepsilon),
\end{equation}
where $g^{(n)}(\varepsilon)$ is an unknown function representing the volume of a neighborhood of the Ising configuration in the phase space of the O$(n)$ model. Since the function $g^{(n)}$ comes from the evaluation of local integrals over a 
neighborhood of the phase space, one expects that it is regular in all the energy density range. 
Equation (\ref{intro:ansatz}) will be discussed in detail in Sec.\ \ref{sec1:sec_dos}. The assumptions made in \cite{prl2011} to derive Eq.\ (\ref{intro:ansatz}) are rather crude and uncontrolled. However, were this relation exact, there would be an interesting consequence: the critical energy densities of the phase transitions of all the O$(n)$ models on a given lattice would be the same and equal to that of the corresponding Ising model. 

Despite the fact that Eq.\ (\ref{intro:ansatz}) is approximate\footnote{In this form, Eq.\ (\ref{intro:ansatz}) cannot be exact neither for finite systems nor in the thermodynamic limit \cite{jstat2012} because it would imply wrong values of the specific heat critical exponent $\alpha^{(n)}$, that would be given by $\alpha^{(n)}=-\alpha^{(1)}$ for any $n$. 
However, it correctly reproduces the sign of the critical exponent that entails a cusp-like behavior of the specific heat at criticality rather than a divergence, as happens in the Ising case.} the critical energy densities are indeed, if not equal, very close to each other, whenever a phase transitions is known to take place, at least for ferromagnetic models defined on regular $d-$dimensional hypercubic lattices. In particular, the critical 
energy densities are the same and equal to the Ising one for all the O$(n)$ models with long-range interactions, as shown by the exact solution \cite{CampaGiansantiMoroni:jpa2003}, and the same happens for all the O$(n)$ models on an one-dimensional lattice 
with nearest-neighbor interactions. For what concerns O$(n)$ models with nearest-neighbor interactions defined on (hyper)cubic lattices with $d>1$, only numerical results are available for the critical energy densities. In $d=2$ a Bere\v{z}inskij-Kosterlitz-Thouless (BKT) phase transition is present in the $n=2$ case, occurring at a value of the energy density that differs of about $2\%$ from the Ising case (see \cite{prl2011,jstat2012} for a deeper discussion\footnote{As to a possible relation between the BKT transition and the 2-$d$ Ising transition, it is interesting to notice that they do share some ``weak'' universality as defined by Suzuki \cite{Suzuki:ptp1974}, despite being very different transitions. The critical exponent ratio $\beta/\nu$ and the exponent $\delta$, for instance, are equal in the two universality classes \cite{Archambault_etal:jpa1997}.}).  
In $d=3$, the difference between the critical energy densities of the O$(n)$ models and those of the Ising model is smaller than $3\%$ for any $n$, including the $n=\infty$ case, as recently shown in \cite{numericalpaper}. In the most frequently considered cases, that is for $n=2$ (the $XY$ model), $n=3$ (the Heisenberg model) and for the O$(4)$ model, it becomes less than $1\%$.

In two particular cases, where the prediction on the critical energy densities is exact, the derivation of Eq.\ (\ref{intro:ansatz}) can be followed rigorously, that is for the mean-field and for the $1d$ ferromagnetic $XY$ models. In these cases, an expression very similar to Eq.\ (\ref{intro:ansatz}), and which reduces to the latter when $\varepsilon\rightarrow\varepsilon_{c}$, can be derived exactly in the thermodynamic limit \cite{jstat2012}. The technical aspects of the derivation strongly rely on the peculiarities of the two models and especially on the fact that they are exactly solvable in the microcanonical ensemble. This feature is crucial for the analysis presented in \cite{jstat2012} and its generalization to O$(n)$ models with $n>2$ in $d>1$ is unlikely to be feasible.

The aim of the present paper is to show that an approximate expression for the density of 
states $\omega^{(n)}$ of nearest-neighbor ferromagnetic O$(n)$ models in $d>1$ can be derived by applying the same energy landscape considerations as discussed above. More precisely, two approximation techniques are presented here: 
the first one can be seen as the natural generalization of the techniques applied in \cite{jstat2012} for the mean-field and for the 1-$d$ XY models and will be referred to as ``first-principle'' approximation in the following; the second one will be named ``ansatz-based'' approximation, 
its starting point being the ansatz on the form of the density of states given by Eq.\ (\ref{intro:ansatz}) 
supposed to be valid in all the energy density range. These techniques can be applied to estimate the density 
of states of in principle any O$(n)$ model in $d>1$. However we have explicitly performed the calculations only for $n=2$, i.e., for the $XY$ model in $d=2$; the technical aspects of the generalization to other members of the O$(n)$ class in $d>2$ 
will be discussed underway.

The paper is organized as follows. In Sec.\ \ref{sec1:sec_dos} the stationary-point approach and the approximations introduced in \cite{prl2011} leading to Eq.\ (\ref{intro:ansatz}) will be recalled and discussed. In Sec.\ \ref{sec2:2dXYstudy} the ``first-principle'' and the ``ansatz-based'' approaches will be introduced. In both approaches a crucial point is the estimation of $g^{(n)}$. A possible way in which this 
can be done will be presented in Sec.\ \ref{SecLMFA}.
In Secs.\ \ref{FPA} and \ref{ansatz} the ``first-principle'' and the ``ansatz-based'' approximations will be, respectively, discussed in details through their application to the $XY$ model in $d=2$. Some remarks on the technical limits of the ``first-principle'' approach will be included in Sec.\ \ref{subsec:degeneracyfactor}. The paper ends with some concluding remarks in Sec.\ \ref{ch5fine}.

\section{Stationary points and density of states}\label{sec1:sec_dos}
In our analysis we are going to consider some special cases of classical O$(n)$ spin models defined on a regular square lattice in $d=2$ and with 
periodic boundary conditions. In these models, to each lattice site $i$ an $n$-component classical 
spin vector $\mathbf{S}_i = (S_i^1,\ldots,S_i^n)$ of unit length is assigned. 
The energy of the model is given by the Hamiltonian
\begin{equation}\label{H-On}
H^{(n)} = - J \sum_{\langle i,j \rangle} \mathbf{S}_i \cdot \mathbf{S}_j= 
- J \sum_{\langle i,j \rangle} \sum_{a = 1}^n S^a_i S^a_j,
\end{equation}
where the angular brackets denote a sum over all pairs of nearest-neighbor lattice sites. 
The exchange coupling $J$ will be assumed positive, resulting in ferromagnetic 
interactions, and without loss of generality we shall set $J=1$ in the following. 
The Hamiltonian (\ref{H-On}) is globally invariant under the $O(n)$ group. In the case $n=1$, 
the symmetry group $O(1)\equiv\mathbb{Z}_{2}$ is a discrete one and the Hamiltonian (\ref{H-On}) becomes the 
Ising Hamiltonian 
\begin{equation}\label{sec1:H-Ising}
 H^{(1)}=-\sum_{\langle i,j\rangle}\sigma_{i}\sigma_{j},
\end{equation}
where $\sigma_{i}=\pm1\;\forall i$. In all the other cases $n\geq2$, the $O(n)$ group is a continuous one. For $n=2$ we obtain the $XY$ model that will be the subject of our study. In this model the spins are constrained on the unit circle $\mathbb{S}^{(1)}$ and the components of the $i$th spin can be 
parametrized by a single angular variable $\vartheta_{i}\in[0,2\pi)$ such that
\begin{equation}\label{XY_components}
\begin{cases}
S^1_i & = \cos\vartheta_i\, ,\\
S^2_i & = \sin\vartheta_i\, .
\end{cases}
\end{equation}
The Hamiltonian of the $XY$ model can thus be conveniently written as 
\begin{equation}\label{sec1:H-XY}
H^{(2)} = - \frac{1}{2}\sum_{i = 1}^N \sum_{j\in\mathcal{N}(i)} \cos\left(\vartheta_i - \vartheta_j \right),
\end{equation}
where $\mathcal{N}(i)$ denotes the set of nearest neighbors of lattice site $i$. 
The energy density $\varepsilon = H^{(2)}/N$ lies in the energy range $[-d,d]$ where $d$ is the lattice dimension. 

Let us now recall the derivation of Eq.\ (\ref{intro:ansatz}) made in \cite{prl2011}. The stationary configurations of $H^{(n)}$ for $n\geq2$ are given by the solutions 
$\bar{S}=(\bar{S}_{1},\ldots,\bar{S}_{N})$ of the $N$ vector equations
\begin{equation}\label{sec1:stat-config1}
 \nabla H^{(n)}=0,
\end{equation}
such that the constraint $\sum_{a=1}{n}\left(S^{a}_{i}\right)^{2}=1$ is satisfied $\forall i=1,\ldots,N$. To find explicitly 
all the stationary configurations is an essentially impossible task but in some special cases, see e.g.\ \cite{MehtaKastner:annphys2011}, or for very small systems, see e.g.\ \cite{Mehta:prerap2011} and references therein. 
However, as shown in Ref.\ \cite{prl2011}, a particular class of solutions can be found, given by all the configurations in which the spins are parallel or anti-parallel to a fixed direction (say the $n$th direction): 
$S^{1}_{i}=\ldots=S^{n-1}_{0}=0\;\forall i$. Indeed, in this case, 
the constraint $\sum_{a=1}{n}\left(S^{a}_{i}\right)^{2}=1$ can be fulfilled by choosing 
$S^{n}_{i}=\sigma_{i}\,\forall i$ and the stationary points equations (\ref{sec1:stat-config1}) are satisfied by any of the 
$2^{N}$ possible choices of the $\sigma$'s. The Hamiltonian (\ref{H-On}) becomes the Ising Hamiltonian (\ref{sec1:H-Ising}) 
on this class of stationary configurations. Therefore a one-to-one correspondence between a class of stationary configurations of the 
Hamiltonian (\ref{H-On}) and \textit{all} the configurations of the Ising model\footnote{These considerations are valid also for 
O$(n)$ models defined on generic graphs and with a 
generic interaction matrix $J_{ij}$ in Eq.\ (\ref{H-On}).}; the corresponding stationary values are the energy 
levels of the Ising Hamiltonian. We shall refer to the class of 
stationary configurations $\bar{S}=(0,\ldots,0,\sigma_{i})\;\forall i=1,\ldots,N$ as ``Ising stationary configurations''. 
Although this class does not include all the stationary points of $H^{(n)}$, see e.g.\ \cite{pre2013}, the $2^N$ Ising ones are a non-negligible fraction of the whole, especially for large $N$. The total number of stationary configurations is expected to be $\mathcal{O}(e^{N})$ \cite{MehtaKastner:annphys2011,Schilling:physicad2006}. When ferromagnetic O$(n)$ models defined on regular $d-$dimensional lattices are considered, as in our case, in the thermodynamic limit $N\rightarrow\infty$ the energy density 
levels of the Ising Hamiltonian (\ref{sec1:H-Ising}) become dense and cover the whole energy density range of all the O$(n)$ 
models. This latter observation, together with the above mentioned properties of the Ising points, suggests that Ising 
stationary configurations are the most important ones, so that the density of states $\omega^{(n)}$ 
of an O$(n)$ model may be approximated in terms of these configurations.

Let us then rewrite the density of states $\omega^{(n)}$ as a sum of integrals over a partition of the phase 
space, 
\begin{equation}\label{sec1:coarea_part2}
\omega^{(n)}(\varepsilon) = \sum_p \int_{U_{p}} \delta(H^{(n)} - N\varepsilon) \, d\Gamma\, , 
\end{equation}
where $p$ runs over the $2^N$ Ising stationary configurations, $U_p$ is a neighborhood of the $p$-th Ising 
configuration such that $\left\{U_p\right\}_{p=1}^{2^N}$ is a proper partition of the configuration space $\Gamma_{N}$, 
that coincides with phase space for spin models defined by the Hamiltonian $H^{(n)}$ in Eq.\ (\ref{H-On}). 

Two approximations were introduced in \cite{prl2011} to derive 
Eq.\ (\ref{intro:ansatz}) from Eq.\ (\ref{sec1:coarea_part2}): 
(\textit{i}) it was assumed that the integrals in Eq.\ (\ref{sec1:coarea_part2}) depend only 
on $\varepsilon$, i.e., the neighborhoods $U_{p}$ can be chosen, or deformed, such as
\begin{equation}\label{g2}
g^{(n)}(\varepsilon,p)=\int_{U_{p}} \delta(H^{(n)} - N\varepsilon) \, d\Gamma  
= g^{(n)}(\varepsilon,q)=\int_{U_{q}} \delta(H^{(n)} - N\varepsilon) \, d\Gamma 
= g^{(n)}(\varepsilon) \,, 
\end{equation}
for any $p,q$ such that $H^{(n)}(p) = H^{(n)}(q) = N\varepsilon$; 
(\textit{ii}) Since non-Ising stationary configurations have been neglected in this analysis, only neighborhoods centered around stationary configurations at energy density $\varepsilon$ have been retained in the sum (\ref{sec1:coarea_part2}). These two assumptions immediately lead to Eq.\ (\ref{intro:ansatz}), that we rewrite for convenience:
\begin{equation}\label{sec1:omega_appr_richiamo}
\omega^{(n)}(\varepsilon) \approx \omega^{(1)}(\varepsilon) \,g^{(n)}(\varepsilon)\, . 
\end{equation}
Both assumptions are needed to derive Eq.\ (\ref{sec1:omega_appr_richiamo}), and are strictly related to each other. However, these two assumptions might well play a very different r\^ole. In \cite{jstat2012} it has been shown 
that in the two analytically tractable special cases, the mean-field and the one-dimensional $XY$ models, assumption (\textit{ii}) does not hold in general: 
it holds only when $\varepsilon \to \varepsilon^{(n)}_c$. As a consequence, one should include also stationary configurations with energy $\varepsilon' \not = \varepsilon$ in the sum. By introducing the continuous function 
\begin{equation}\label{sec1:bigG}
G^{(n)}(\varepsilon,\varepsilon^{\prime})=\int_{U_{p}|\frac{H^{(n)}(p)}{N}=\varepsilon^{\prime}}
\delta\left(H^{(n)}-N\varepsilon\right)d\Gamma  
\end{equation}
such that 
\begin{equation}\label{sec1:bigG-vs-smallg}
G^{(n)}(\varepsilon,\varepsilon)=g^{(n)}(\varepsilon)\,,
\end{equation}
the density of states can be written as
\begin{equation}\label{sec1:omega_prod}
\omega^{(n)}(\varepsilon) = \omega^{(1)}(\tilde{\varepsilon}) \, G^{(n)}(\varepsilon,\tilde{\varepsilon})\,.
\end{equation}
In the above expression $\tilde{\varepsilon}$ is a suitable function of $\varepsilon$. If 
$\tilde{\varepsilon} = \varepsilon$, then using Eq.\ (\ref{sec1:bigG-vs-smallg}) one recovers Eq.\ (\ref{sec1:omega_appr_richiamo}). This precisely happens in the mean-field and 1-$d$ $XY$ models when $\varepsilon\rightarrow\varepsilon^{(n)}_{c}$, that is at the phase transition. 

To compute $G^{(n)}(\varepsilon,\tilde{\varepsilon})$ the microcanonical solutions of the models are needed, as shown in 
\cite{jstat2012}. For this reason Eq.\ (\ref{sec1:omega_prod}) becomes of little use when O$(n)$ models in $d>1$ are considered, as in our case. Therefore in the ``ansatz-based'' approximation we shall assume that Eq.\ (\ref{sec1:omega_appr_richiamo}) is valid in the whole energy density range and not only for 
$\varepsilon=\varepsilon^{(n)}_{c}$, as we are going to discuss in detail in the following Sections. 

\section{First-principle and ``ansatz-based'' approximations: an overview}\label{sec2:2dXYstudy}
Let us now discuss how the results recalled in Sec.\ \ref{sec1:sec_dos} can be generalized to O$(n)$ spin models in $d>1$. Our results can be grouped into two different categories according to the approach involved in their derivation.

The first attempt in the generalization is to set up
an approximation procedure that allows to estimate each integral appearing in the sum in
Eq.\ (\ref{sec1:coarea_part2}), that yields the density of states $\omega^{(n)}(\varepsilon)$. This approach is the direct generalization of the 
techniques already applied to the mean-field and to the one-dimensional $XY$ models to generic short-range O$(n)$ models except that, for short range systems, only an approximate evaluation of each of these integrals is possible.
This approximation protocol will be denoted as ``first-principles'' approximation, its starting 
point being simply the density of states $\omega^{(n)}$ expressed in terms of a suitable partition of the 
phase space $\Gamma$ of the system. The procedure will be presented in 
Sec.\ \ref{FPA} and the two-dimensional $XY$ model will be considered as a test model of our analysis.

The second approach starts from the ansatz on the form of the density of states given by 
Eq.\ (\ref{sec1:omega_appr_richiamo}) with $g^{(n)}(\varepsilon)$ as in Eq.\ (\ref{g2}). 
In this approach it is assumed that the integral in Eq. (\ref{g2}) does not depend on the specific point considered but only on its energy density $\varepsilon$ and that only Ising points with energy density $\varepsilon^{\prime}=\varepsilon$ contribute to the density of states $\omega^{(n)}(\varepsilon)$ in Eq.\ (\ref{sec1:omega_appr_richiamo}). We will denote this kind of analysis as ``ansatz-based'' approximation. 
The general aspects of the method will be presented in Sec. \ref{ansatz} and its application to the $XY$ model in two dimensions will be discussed in detail.

Both in the ``ansatz-based'' and  ``first-principle'' approximations, the main point is to estimate the quantity
\begin{equation}\label{continuous-factor}
g^{(n)}(\varepsilon,p)=\int_{U_{p}}d\Gamma\;\delta\left(H^{(n)}-N\varepsilon\right)=
\int_{U_{p}}d\Gamma\,\delta\left(-\frac{1}{2}\sum^{N}_{i=1}\sum_{j\in\mathcal{N}(i)}\sum^{n}_{a=1}
S^{a}_{i}S^{a}_{j}-N\varepsilon\right)\,,
\end{equation}
where the expression of $H^{(n)}$ given in (\ref{H-On}) has been explicitly introduced.
Among the O$(n)$ class of models, Eq.\ (\ref{continuous-factor}) can be analytically evaluated only in the cases of the mean-field and one-dimensional $XY$ models; its computation in the general case being as difficult as to find an 
exact solution of the models. 
However some computational procedures can be set up to carry on the calculations, albeit approximate. 
In \cite{Rachele:thesis} an approximation scheme has been introduced, 
named the Local-Mean-Field (LMF) approximation. This is a model-dependent procedure
that allows to reduce the $N-$dimensional integral in Eq.\ (\ref{continuous-factor}) over the configuration space $\Gamma$ to $N$ one-dimensional integrals over uncoupled variables. 
The uncoupling procedure becomes possible once suitable model-dependent collective variables are defined\footnote{Their r\^{o}le reminds the one played by 
$N_{\pi}$ and $N_{d}$ in \cite{jstat2012} for the mean-field and the one-dimensional $XY$ models, respectively}. 
The LMF procedure will be presented in Sec.\ \ref{SecLMFA}. 
We shall restrict the presentation to the case $n=2$ and to two-dimensional square lattices with periodic boundary conditions. The generalization to higher-dimensional lattices and different values of $n$ should be possible along similar 
lines, but we did not work this out in detail although something similar (but in the canonical ensemble) has been done in \cite{Rachele:thesis} for 
the case $n=2$ and $d=3$. 

Other schemes than LMF for approximating the $g^{(n)}$'s may be used. However, their implementation may be quite complicated in the case of the ``first-principle'' approximation, while feasible in the ``ansatz-based'' one. The reason for this will become clear in Sec.\ \ref{subsec:degeneracyfactor}. We will give an example of an alternative scheme, corresponding to a harmonic approximation of the Hamiltonian around each Ising stationary point, while discussing the ``ansatz-based'' approximation in Sec.\ \ref{ansatz}.

\subsection{The Local Mean-Field (LMF) approximation}\label{SecLMFA}
\begin{figure}
\centering
\includegraphics[width=4cm]{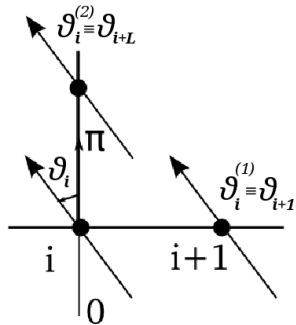} 
\caption{Index convention for the nearest-neigbors spins $\vartheta^{(1)}_{i}$ and 
$\vartheta^{(2)}_{j}$ of $\vartheta_{i}$ for the $XY$ model on a square lattice with edge $L=\sqrt{N}$ and periodic 
boundary conditions. The spins in the lattice sites $i+1$ and $i+L$ are said to be second neighbors.}
\label{nn-convention}
\end{figure}

In the case $n=2$ and $d=2$, i.e. for the $XY$ model in two spatial dimensions, the parametrization (\ref{XY_components}) can be chosen such that the integral in Eq. (\ref{continuous-factor}) becomes
\begin{equation}\label{continuous-factor-XY}
 \begin{split}
&\int_{U_{p}}d\vartheta_{1}\ldots d\vartheta_{N}\,{\delta\left[-\frac{1}{2}\sum^{N}_{\langle i,j\rangle} 
\cos\left(\vartheta_{i}-\vartheta_{j}\right)-N\varepsilon\right]}=\\
=&\int_{U_{p}}d\vartheta_{1} \ldots d\vartheta_{N}
\delta {\left[-\sum^{N}_{i=1}\left(\cos\vartheta_{i} \sum^{2}_{j=1}
\cos\vartheta^{(j)}_{i}+\sin\vartheta_{i} \sum^{2}_{j=1}\sin\vartheta^{(j)}_{i}\right)-N\varepsilon\right]}=\\
=&\int_{-\infty}^{\infty} \frac{dk}{2\pi}e^{-ikN\varepsilon}
\int_{U_{p}}d\vartheta_{1} \ldots d\vartheta_{N}\;
e^{-i k \sum^{N}_{i=1} \left[ \cos\vartheta_{i} \sum^{2}_{j=1}\cos \vartheta^{(j)}_{i} + \sin\vartheta_{i} \sum^{2}_{j=1}
\sin\vartheta^{(j)}_{i} \right]}
\end{split}
\end{equation}
where we have introduced the integral representation of the $\delta-$function, 
$\delta(x)=\frac{1}{2\pi}\int_{-\infty}^{\infty}e^{-ikx}dk$; 
$\vartheta^{(j)}_{i}$ denotes the nearest-neighbor spin $j$ of the lattice site $i$ considered according to 
the convention reported in Fig.\ \ref{nn-convention}. 
The two spins identified by $\vartheta^{(1)}_{i}$ and $\vartheta^{(2)}_{i}$ are said to be second-neighbor spins.
A generalization of Eq.\ (\ref{continuous-factor-XY}) to $n>2$ and $d>2$ should be straightforward once a proper 
parametrization for the angular variables as in Eq. (\ref{XY_components}) is chosen and the spins are labeled accordingly. In Eq.\ (\ref{continuous-factor-XY}), $p$ denotes any Ising stationary point; the LMF approximation can then be introduced as 
follows.

Let us consider a particular Ising point, i.e., 
$\bar{p}=\left(\bar{\vartheta}_{1},\ldots,\bar{\vartheta}_{N}\right)$ with 
$\bar{\vartheta}_{i}\in\{0,\pi\}\;\forall i=1,\ldots,N$. For each angular variable $\vartheta_{i}$, 
in Eq.\ (\ref{continuous-factor-XY}) we replace the $N-1$ variables $\vartheta_{j}$ with $j\neq i$ 
with their values $\bar{\vartheta}_{j}$ at point $\bar{p}$. 
The variable $\vartheta_{i}$ is left free to vary in all the range specified by $U_{\bar{p}(\vartheta_{i})}$. 
In this way the angular variables in Eq.\ (\ref{continuous-factor-XY}) become uncoupled and 
\begin{equation}\label{LMFA}
\begin{split}
&\int_{-\infty}^{\infty} \frac{dk}{2\pi}e^{-ikN\varepsilon}
\int_{U_{p}}d\vartheta_{1} \ldots d\vartheta_{N}\;e^{-i k \sum^{N}_{i=1} \left[ \cos\vartheta_{i} 
\sum^{2}_{j=1}\cos \vartheta^{(j)}_{i} + \sin\vartheta_{i} \sum^{2}_{j=1}\sin\vartheta^{(j)}_{i} \right]}=\\
&=\int_{-\infty}^{\infty}\frac{dk}{2\pi} 
e^{-ikN\varepsilon}\;\prod^{N}_{i=1}\int_{U_{\bar{p}(\vartheta_{i})}} d\vartheta_{i}\, 
e^{-ik \cos\vartheta_{i} \left(\cos\bar{\vartheta}^{(1)}_{i}+\cos\bar{\vartheta}^{(2)}_{i}\right)} 
\end{split}
\end{equation}
with $H(\bar{p})=N\varepsilon^{\prime}$. 
Eq.\ (\ref{LMFA}) clarifies the choice of ``Local Mean-Field'' as the name for the approximation. 
Indeed, after LMF approximation is applied, the angular variables $\vartheta_{i}$'s in Eq. (\ref{LMFA}) become independent variables in $U_{\bar{p}(\vartheta_{i})}$; 
each degree of freedom interacts only with a sort of local mean-field generated by the spins located in the nearest-neighbor lattice sites. 
Remarkably, with this approximation the contribution of the sines in the exponent of Eq.\  (\ref{LMFA}) vanishes, since $\sin\vartheta_{i}=0$ when $\vartheta_{i}\in\{0,\pi\}$, and the expression in Eq.\ (\ref{LMFA}) simplifies. 

Since the analysis is restricted to Ising stationary configurations, in Eq.\ (\ref{LMFA}) only two cases are possible:
\begin{itemize}
 \item [(a)] The second-neighbor spins are equal. 
 In this case $\cos \bar{\vartheta}^{(1)}_{i} = \cos\bar{\vartheta}^{(2)}_{i}=\pm 1$, and so
\begin{equation}\label{LMFAfact1}
\int_{U_{\bar{p}(\vartheta_{i})}} d\vartheta_{i}\;e^{-ik \cos\vartheta_{i} \left(\cos\bar{\vartheta}^{(1)}_{i}+
\cos\bar{\vartheta}^{(2)}_{i}\right)}=\int_{U_{\bar{p}(\vartheta_{i})}}d\vartheta_{i}\;e^{\mp2ik\cos\vartheta_{i}};
\end{equation}
\item [(b)] The second-neighbor spins are opposite.
In this case $\cos\bar{\vartheta}^{(1)}_{i}=-\cos\bar{\vartheta}^{(2)}_{i}$, and so
\begin{equation}\label{LMFAfact2}
\int_{U_{\bar{p}(\vartheta_{i})}}d\vartheta_{i}\;e^{-ik\cos\vartheta_{i}
\left(\cos\bar{\vartheta}^{(1)}_{i}+\cos\bar{\vartheta}^{(2)}_{i}\right)}=
\int_{U_{\bar{p}(\vartheta_{i})}}d\vartheta_{i}\;.
\end{equation}
\end{itemize}
To evaluate Eqs.\ (\ref{LMFAfact1}) and (\ref{LMFAfact2}), the 
neighborhoods $U_{\bar{p}(\vartheta_{i})}$ have to be defined. 
A priori there are no particular clues on the best choice of $U_{\bar{p}(\vartheta_{i})}$, 
the only request being that $\{U_{p(\vartheta_{1})},\ldots,U_{p(\vartheta_{N})}\}^{2^{N}}_{p=1}$ has to be a partition of 
the phase space of the system. In the following, two different choices of $U_{p(\vartheta_{i})}$ will be considered:
\begin{equation}\label{UpChoice1}
U_{p(\vartheta_{i})}=\begin{cases}\left[-\pi,\pi\right],\; &\mbox{if $\bar{\vartheta}_{i}=0$,}\\
\left[0,2\pi\right],\; &\mbox{if $\bar{\vartheta}_{i}=\pi$.}
\end{cases}
\end{equation}
\begin{equation}\label{UpChoice2}
U_{p(\vartheta_{i})}=\begin{cases}\left[-\frac{\pi}{2},\frac{\pi}{2}\right],\; &\mbox{if $\bar{\vartheta}_{i}=0$,}\\
\left[\frac{\pi}{2},\frac{3\pi}{2}\right],\; &\mbox{if $\bar{\vartheta}_{i}=\pi$.}
\end{cases}
\end{equation}
In principle the choice in Eq.\ \eqref{UpChoice1} should be avoided, the neighborhoods $U_{p(\vartheta_{i})}$'s being not disjoint. We will anyhow consider it in the following, since it is the easiest choice that can be done a priori for these systems. For 
the ``first-principle'' approximation only the first choice for the $U_{p(\vartheta_{i})}$ will be considered. The technical reasons for this choice will be discussed in Sec. \ref{subsec:degeneracyfactor}. 
For the ``ansatz-based'' approximations, instead, both \eqref{UpChoice1} and \eqref{UpChoice2} will be considered and the effect of neighborhood superposition will be explicitly discussed.

\section{``First-principle'' approximation}\label{FPA}
Let us consider the form of the density of states $\omega^{(n)}(\varepsilon)$ given by Eq.\ (\ref{sec1:coarea_part2}). 
In the following we are going to apply our procedure to derive an approximate form of the density of states 
$\omega^{(2)}(\varepsilon)$ for the $XY$ model in $d=2$. A ge\-ne\-ra\-li\-za\-tion of these techniques 
to O$(n)$ models with $n>2$ in $d>2$ is thought to be possible on the same lines and the key points will be highlighted 
in the following discussion. 

Once the $XY$ model in $d=2$ is considered, Eq.\ (\ref{sec1:coarea_part2}) can be written as 
\begin{equation}\label{FPAstep1}
\omega^{(2)}_{N}(\varepsilon)\simeq \sum_{p\in\Gamma}\frac{1}{2\pi}\int_{-\infty}^{\infty}dk\, 
e^{-iNk\varepsilon}\; \int_{U_{p}}d\vartheta_{1} \ldots d\vartheta_{N} e^{-i k \sum^{N}_{i=1} \left[ \cos\vartheta_{i} 
\sum^{2}_{j=1}\cos \vartheta^{(j)}_{i} + \sin\vartheta_{i} \sum^{2}_{j=1}\sin\vartheta^{(j)}_{i} \right]},
\end{equation}
where $p$ is any Ising stationary configuration. 
The integral over the angular variables $\vartheta_{i}$ can be evaluated by applying the LMF approximation presented in Sec.\ \ref{SecLMFA}. In this case we choose a definition of the integration neighborhoods as in Eq.\ (\ref{UpChoice1}). 
Similar results are supposed to hold also for a choice of 
$U_{p(\vartheta_{i})}$ as in Eq.\ (\ref{UpChoice2}) but the 
calculations have not been carried out in this case. The generalization 
of Eq. (\ref{FPAstep1}) to other O$(n)$ models would depend on the parametrization chosen to describe the spin variables and on the number of nearest neighbors (that is, on the dimensionality of the lattice). 

From Eqs.\ (\ref{LMFAfact1}) and (\ref{LMFAfact2}) we have
\begin{equation}\label{LMFAUp1a}
\int_{-\pi}^{\pi} d\vartheta \;e^{\mp2ik\cos\vartheta}=2\pi J_{0}(2|k|)
\end{equation}
whenever a couple of equal second-neighbors spins is present in the system, and
\begin{equation}\label{LMFAUp1b}
\int_{0}^{2\pi} d\vartheta \;1=2\pi
\end{equation}
whenever a couple of opposite second-neighbors spins is present in the system; 
$J_{0}(x)$ is the zero-order Bessel function of the first kind \cite{AbramowitzStegun:book}. 
We will denote by $N_{c}$ the number of couples of equal second-neighbor spins\footnote{The number of couples of opposite spins will be simply given by $N_{D}=N-N_{c}$.} and by $n_{c}=N_{c}/N$ its number density. 

For any given value of $N_{c}$, a particular family of Ising stationary configurations is selected and the integral in Eq.\ (\ref{FPAstep1}) becomes the product of two 
different contributions: $\left(2\pi J_{0}(2|k|)\right)^{N_c}$ due to couples of 
equal second neighbors spins, 
and $\left(2\pi\right)^{N-N_{c}}$ due to the opposite couples. 
In this way Eq.\ (\ref{FPAstep1}) becomes
\begin{equation}\label{FPAstep2}
\omega^{(2)}_{N}(\varepsilon)\simeq\sum_{N_{c}=0}^{N}\;\nu(N_{c})\frac{(2\pi)^{N(1-n_{c})}}{2\pi} 
\int_{-\infty}^{\infty}dk\;e^{N\left[-i\, k\, \,\varepsilon +\,n_{c}\log\left(2\pi J_{0}(2\sqrt{k^{2}})\right) \right]},
\end{equation}
where $\nu(N_{c})$ is a degeneracy factor counting the number of Ising configurations with a given value $N_{c}$ of the collective variable. 
Due to periodic boundary conditions, determining $\nu(N_{c})$ in Eq.\ (\ref{FPAstep2}) reduces to a combinatorial problem analogous to disposing $N_{c}$ distinct elements over $N$ possible empty spaces; we then have
$\nu(N_{c})=\binom {N}{N_{c}}=\frac{N!}{N_{c}!(N-N_{c})!}$. The evaluation of the degeneracy factor $\nu(N_{c})$ is crucial 
for this kind of analyses and we will come back on this point in Sec.\ \ref{subsec:degeneracyfactor}.

\begin{figure}
\centering
\includegraphics[width=10cm]{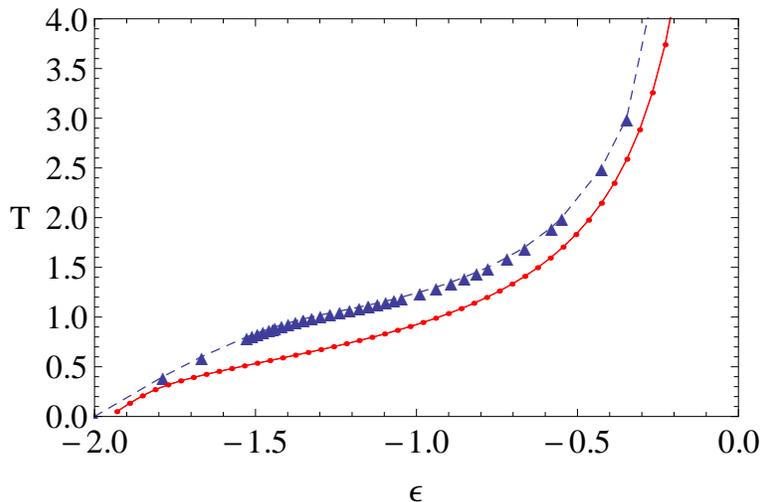} 
\caption{Temperature $T$ as a function of the energy density 
$\varepsilon$ as derived from Eq. (\ref{FPAstep4}) for the $XY$ model in $d=2$. 
Our results (red circles) are plotted together with the data obtained by a Monte Carlo simulation (blue triangles). Errorbars of the simulation data are smaller than symbol sizes.
The dashed blue line and the solid red line connecting the two sets of data are meant as guide to the eyes.}
\label{binomialEnergy}
\end{figure}

\begin{figure}
\begin{center}
\includegraphics[width=10cm]{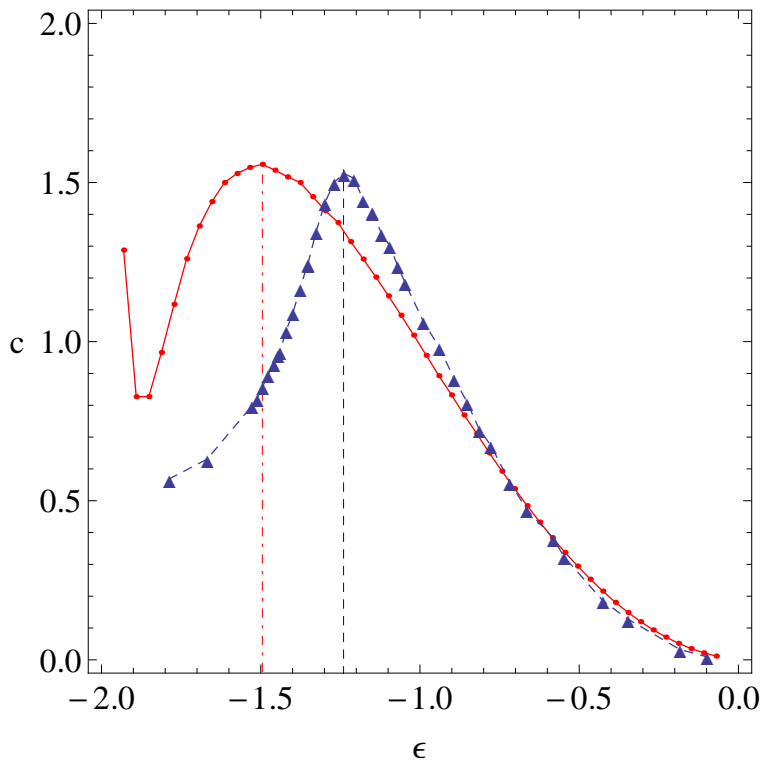}
\caption{Specific heat $c$ as a function of the energy density 
$\varepsilon$ as derived from Eq. (\ref{FPAstep4}) for the $XY$ model in $d=2$. 
Our results (red circles) are plotted together with the data obtained by a Monte Carlo simulation (blue triangles). Errorbars of the simulation data are smaller than symbol sizes.
The dashed blue line and the solid red line connecting the two sets of data are meant as guide to the eyes.}
\vspace{-0.1cm}
\label{BinomialCv}
\end{center}
\end{figure}

Since we are interested in the large-$N$ behavior of the system, the integration over $k$ in Eq.\ (\ref{FPAstep2}) can be computed using the saddle-point method; the saddle-point equation is given by $k=i\tau$ with $\tau$ satisfying the self-consistency equation
\begin{equation}\label{tauSC}
\frac{I_{1}(2\tau)}{I_{0}(2\tau)}=\frac{\varepsilon}{2n_{c}}.
\end{equation}
Eq.\ (\ref{FPAstep2}) can then be written as
\begin{equation}\label{FPAstep3}
\begin{split}
\omega^{(2)}_{N}(\varepsilon)\simeq&\sum_{Nn_{c}=0}^{N}\;\frac{N!}{N_{c}!(N-N_{c})!}\,
e^{N\left[-\tau\,\varepsilon+\,\log2\pi+\,n_{c}\log\left(I_{0}(2\tau)\right)\right]}\\
\simeq&\;N\,\int_{0}^{1}dn_{c}\;e^{N\left[-\tau\,\varepsilon-\,n_{c}\log n_{c}-(1-n_{c})\log(1-n_{c})+
\log2\pi+\,n_{c}\log\left(I_{0}(2\tau)\right)\right]},
\end{split}
\end{equation}
where we have replaced the sum over $N_{c}$ with an integration over $n_{c}$, we have neglected the term $-\frac{1}{N}\log{2\pi}$ in the exponent since its contribution vanishes for $N\to\infty$, and we have introduced the Stirling approximation of the factorial terms in the binomial coefficient to get
\begin{equation}
\frac{N!}{N_{c}!(N-N_{c})!}\simeq e^{N\left[-n_{c}\log n_{c}-(1-n_{c})\log(1-n_{c})\right]}.
\end{equation}
For each value of the energy density $\varepsilon$, we can approximate Eq.\ (\ref{FPAstep3}) as
\begin{equation}
\omega^{(2)}_{N}\simeq \;N\;e^{N\;\max_{n_{c}\in[0,1]}\;
\left[-\tau\,\varepsilon-\,n_{c}\log n_{c}-(1-n_{c})\log(1-n_{c})+\log2\pi+\,n_{c}\log\left(I_{0}(2\tau)\right)\right]}\,.
\end{equation}
The entropy density $s^{(2)}(\varepsilon)$ in the thermodynamic limit is finally given by:
\begin{equation}\label{FPAstep4}
s^{(2)}(\varepsilon)\simeq \max_{n_{c}\in\left[0,1\right]}f(n_{c},\tau),
\end{equation}
with 
\begin{equation}
f(n_{c},\tau)=-\tau\,\varepsilon-\,n_{c}\log n_{c}-(1-n_{c})\log(1-n_{c}) + \\
+\log2\pi+\,n_{c}\log\left(I_{0}(2\tau)\right)
\end{equation}
and $\tau$ numerically determined from Eq.\ (\ref{tauSC}). The maximization procedure in Eq.\ (\ref{FPAstep4}) can be performed numerically. The temperature $T$ and the specific heat $c$ as a function of the energy density can be 
computed from Eq.\ (\ref{FPAstep4}) by numerical differentiation and are shown as red circles in Figs.\ \ref{binomialEnergy} 
and \ref{BinomialCv}, respectively. 
As a comparison, data for $T$ and $c$ as functions of the energy density $\varepsilon$ have been computed with a Monte Carlo simulation of the $XY$ model with edge $L=32$ in $d=2$, performed with the optimized cluster algorithm 
{\tt{spinmc}} provided by the ALPS project \cite{ALPS}. The numerical data are plotted in 
Figs.\ \ref{binomialEnergy} and \ref{BinomialCv} as blue squares together with the results obtained from our approximation. 

Fig.\ \ref{binomialEnergy} shows that Eq.\ (\ref{FPAstep4}) correctly reproduces the asymptotic behavior of the function 
$T(\varepsilon)$ both in the 
low and in the high energy regime, at a semiquantitative level. 
In particular, for $\varepsilon\simeq-2$ our results and the simulation data are almost coincident. 
For $\varepsilon\geq-1.8$ the difference between the numerical and the approximate results increases; the approximate 
value of the temperature remains lower than the results obtained from the simulations although essentially 
at a constant distance. The difference between the calculated and the simulated temperatures never exceeds $15\%$.

Our results for the specific heat are reported in Fig.\ \ref{BinomialCv} (red circles). They show a peak for  $\varepsilon_{p,1}\simeq-1.495$ marked by the vertical red dot-dashed line, at a slightly lower energy density value than $\varepsilon_{p}\simeq -1.24$ 
where the peak occurs in the simulation data (vertical dashed blue line). 
The overall shape of the specific heat sketched by our results is 
in qualitative agreement with the numerical results for $\varepsilon\in[\varepsilon_{p,1},-0.6]$, although shifted to lower energies.  
For $\varepsilon\geq-0.6$ the agreement increases also quantitatively and two sets of points 
become essentially indistinguishable. 
On the other side, for $\varepsilon<\varepsilon_{p,1}$ the agreement becomes worse. 
Both from theoretical and numerical results, 
we know that $c(\varepsilon)\rightarrow0.5$ when $\varepsilon\rightarrow0$, 
see i.e. \cite{KosterlitzThouless:jphysc1973,berezinskij:sovphysjetp1971}. Our results show an abrupt increase for $\varepsilon\simeq-1.85$. This is a shortcoming of our approximation that is still under investigation. 

\subsection{The degeneracy factor $\nu$ and a different approximation for $\omega^{(n)}$}\label{subsec:degeneracyfactor}
The analysis presented in Sec.\ \ref{FPA} shows that the ``first-principle'' procedure provides a practical method to derive an approximate form of the density of states $\omega^{(2)}$ in two dimensions. The thermodynamic functions derived from Eq.\ (\ref{FPAstep4}) are in reasonably good agreement with the 
simulations, as shown in Figs. \ref{binomialEnergy} and \ref{BinomialCv}. This suggests that our idea of considering only Ising stationary points in the derivation 
of $\omega^{(2)}(\varepsilon)$ is trustworthy and provides a good strategy to approximate
the thermodynamic properties of continuous O$(n)$ models, in principle for any value of $n$ and $d$.

An important feature of the ``first-principle'' approximation is the natural emergence of collective variables, like $N_{c}$, in terms of which the stationary configurations can be parametrized and the density of states re-written as in 
Eq. (\ref{FPAstep2}). The number and the type of the collective variables depend on several aspects: the model considered, the dimensionality of the lattice, the definition of the integration neighborhoods $U_{p(\vartheta_{i})}$, the specific Ising 
point $p$ considered, and the approximation strategy applied to evaluate the integral in Eq.\ (\ref{continuous-factor}). 
Indeed, as we are going to show in Sec.\ \ref{sub:harmonic}, 
instead of applying the LMF approximation other strategies could have been adopted 
to compute the integral in Eq.\ (\ref{continuous-factor}), like e.g.\ a harmonic expansion 
of $H^{(n)}$ around the Ising stationary points. 
In this case other quantities, as the determinant $\mathcal{D}(p)$ of the 
Hessian matrix of $H^{(n)}$ or the density of index $\iota(p)=\frac{\mathcal{I}}{N}(p)$ would have emerged in the analysis\footnote{More precisely, the quantity that would appear is $\mathcal{D}(p)^{\frac{1}{N}}$ and will be referred to as the ``reduced determinant'' of the Hessian 
matrix of $H^{(n)}$ in $p$; the index $\mathcal{I}$ of a stationary point $p$ is the number of negative eigenvalues of the Hessian matrix of the Hamiltonian evaluated at $p$.}. 

Let us denote by $\mathbf{z}(n,d,p)$ the vector of collective variables needed in the evaluation. In the ``first-principle'' approach the density of states $\omega^{(n)}_{N}(\varepsilon)$ can always be reduced to a form of the type
\begin{equation}\label{spiegazione1}
 \omega^{(n)}_{N}(\varepsilon)=\sum_{\mathbf{z}(n,d,p)}^{N}\nu(\mathbf{z}(n,d,p),N)\;f(\mathbf{z}(n,d,p),N),
\end{equation}
as shown in Eq.\ (\ref{FPAstep2}). In the above expression $\nu(\mathbf{z}(n,d,p),N)$ is the degeneracy factor associated to the collective vector of parameters $\mathbf{z}$ counting the number of Ising configurations for given values $\mathbf{z}(n,d,p)$ and $N$ 
of the collective variables and of the number of degrees of freedom of the system, respectively.

The degeneracy factor $\nu$ can not be evaluated analytically but in some specific cases like the one discussed in Sec.\ \ref{FPA} and those discussed in \cite{jstat2012}.
In the analysis presented in Sec.\ \ref{FPA} only integration neighborhoods as in 
Eq.\ (\ref{UpChoice1}) have been considered. Indeed, as will be shown in Sec.\ \ref{subsesctionLMFAUp2}, a choice of the integration neighborhoods as in (\ref{UpChoice2}) would have produced the emergence of two different collective variables, $N_{c}$ and $N_{3}$ 
with $N_{3}$ denoting the number of triplets of equal Ising spins. To compute $\omega^{(2)}$, it would then be necessary to estimate the degeneracy factor $\nu(N_{c},N_{3})$ counting the number of Ising configurations with 
given values of $N, N_{c}$ and $N_{3}$. Up to our knowledge, this quantity is not analytically known. 
To carry on the calculation one may then estimate $\nu(\mathbf{z}(n,d,p))$ numerically. This can be done 
for instance by performing a Monte Carlo simulation in which the Ising configurations 
are grouped and counted according to their value of $\mathbf{z}$.  
This method, although possibly correct, would require a strong computational effort and we preferred to leave it to future work. 

On the other hand the ansatz on the form of the density of states $\omega^{(n)}$ 
given by Eq.\ (\ref{sec1:omega_appr_richiamo}) 
is able to reproduce with unexpected accuracy both the emergence of the phase transitions in the O$(n)$ models and even the critical energy density values at which the transitions are located \cite{prl2011}. Then, one may consider Eq.\ (\ref{sec1:omega_appr_richiamo}) as the new starting point to approximate the thermodynamic functions of the O$(n)$ system in the whole energy density range $[-d,d]$. 
In this kind of approach, 
called ``ansatz-based'' approach, the main point remains the estimation of $g^{(n)}(\varepsilon)$; this implies the 
emergence of collective variables $\mathbf{z}(n,p,d)$, as before. 
This notwithstanding, it is now reasonable to assume that, 
given a particular Ising point $p$ with energy 
density $\varepsilon^{\prime}=H^{(n)}(p)/N$, 
the possible values of the collective variables $\mathbf{z}(n,d,p)$ would narrow around a typical value 
$\tilde{\mathbf{z}}(n,d,p)$ when $N\rightarrow\infty$ (see e.g.\ Ref.\ \cite{pre2013}). In this limit the typical value 
$\tilde{\mathbf{z}}(n,d,p)$ would not depend on $p$ anymore but only on its energy density 
$\varepsilon^{\prime}$. We then have
$\tilde{\mathbf{z}}(n,d,p)\simeq\tilde{\mathbf{z}}(n,d,\varepsilon^{\prime})$. Since the ansatz in 
Eq.\ (\ref{sec1:omega_appr_richiamo}) imposes that only stationary points with energy 
density $\varepsilon^{\prime}=\varepsilon$ have to 
be considered in the evaluation of $\omega^{(n)}(\varepsilon)$, we have that
$g^{(n)}(\mathbf{z}(d,p))\rightarrow g^{(n)}(\mathbf{z}(d,\varepsilon))=g^{(n)}(\varepsilon)$ in $d$ dimensions, 
with $\mathbf{z}(d,\varepsilon)$ suitable functions that can be easily estimated by fits of numerical simulation data.  

All these concepts will be clarified in the next Sections where the ``ansatz-based'' approximation will be applied to 
the $XY$ model in $d=2$. 

\section{``Ansatz-based'' approximation}\label{ansatz}
Let us consider the density of states $\omega^{(n)}$ as given by Eq.\ (\ref{sec1:omega_appr_richiamo}); then, our purpose is 
to estimate $g^{(n)}(\varepsilon)$. 
This can be done for instance by applying the LMF approximation 
introduced in Sec.\ \ref{SecLMFA}. 
Let us consider the $XY$ model in $d=2$ as test model of our procedure so that all the technical tools presented in Sec.\ \ref{SecLMFA} can be immediately applied to this case; the generalization to other O$(n)$ models 
should be straightforward.

\subsection{LMF approximation for $g^{(2)}(\varepsilon)$ and $U_{p(\vartheta_{i})}$ given by Eq.\ (\ref{UpChoice1}).}
\label{subsesctionLMFAUp1}
We start considering the LMF approximation with the first choice of the neighborhoods $U_{p(\vartheta_{i})}$ given by 
Eq.\ (\ref{UpChoice1}). In this case the calculations proceed on the same lines as in the case of the ``first-principles'' approximation, the only difference being that only Ising points with energy density $\varepsilon$ are considered and 
the collective variable $\mathbf{z}(2,2,\varepsilon)=N_{c}(\varepsilon)$ does not depend on the specific Ising point $p$ anymore but only on its energy density $\varepsilon$. 

\begin{figure}
\centering
\includegraphics[width=10cm]{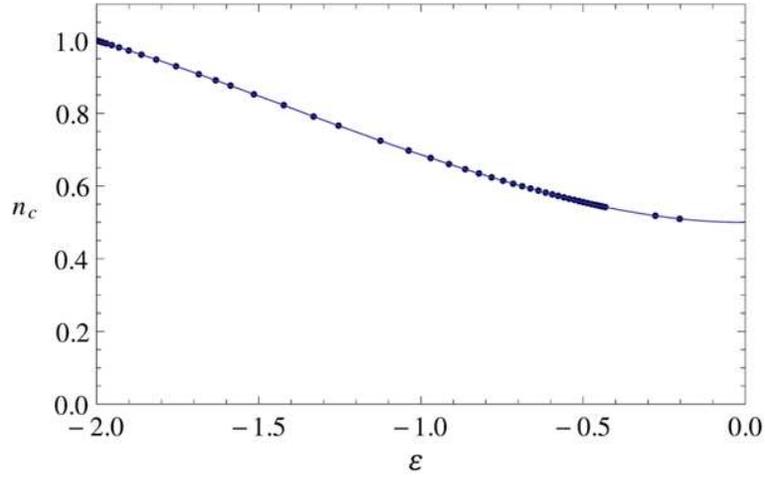} 
\caption{Data for $n_{c}$ obtained by a Monte Carlo simulation of the two-dimensional $XY$ model with edge $L=32$ (blue solid circles). Errorbars of the simulation data are smaller than the symbol sizes. The solid blue line is the interpolating function $n_{c}(\varepsilon)$.}
\label{ncplot}
\end{figure}

The density of states can then be written as
\begin{equation}\label{ansatzUp1omega1}
 \omega^{(2)}_{N}(\varepsilon)=\omega^{(1)}_{N}(\varepsilon)\frac{(2\pi)^{N(1-n_{c}(\varepsilon))}}{2\pi}
 \int_{-\infty}^{\infty}e^{N\left[-ik\varepsilon+n_{c}(\varepsilon)\log\left(2\pi J_{0}\left(2\sqrt{k^2}\right)\right)\right]}
\end{equation}
where $\omega^{(1)}_{N}(\varepsilon)$ plays the r\^{o}le of $\nu(N_{c},N)$ in Eq.\ (\ref{FPAstep2}) and is analytically known thanks to the exact solution of the 2-$d$ Ising model. On the other hand, $n_{c}=n_{c}(\varepsilon)$ is an unknown function that has 
to be determined. This has been done interpolating the numerical data obtained by a 
Monte Carlo simulation of the two-dimensional $XY$ model with edge $L=32$. The result is shown as a solid blue line in Fig.\ \ref{ncplot}, 
together with the simulation data.

\begin{figure}
\centering
\includegraphics[width=10cm]{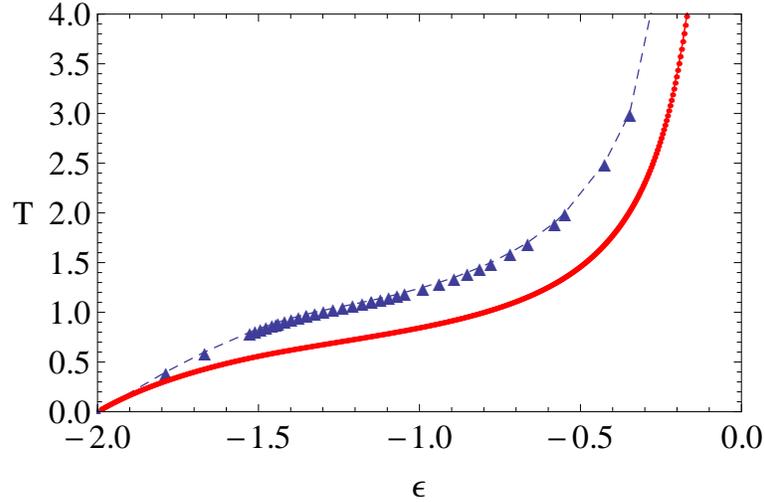} 
\caption{Temperature $T$ as a function of the energy density $\varepsilon$ as derived from Eq.\ (\ref{LMFAUp1sFINALE}) for the $XY$ model in $d=2$. 
Our results (red circles) are plotted together with the data obtained by a Monte Carlo simulation (blue triangles). Errorbars of the simulation data are smaller than the symbol sizes.
The dashed blue line and the solid red line connecting the two sets of data are meant as guide to the eyes.}
\label{LMFAUp1energy}
\end{figure}

The integral in Eq.\ (\ref{ansatzUp1omega1}) can be computed with the saddle-point method and the saddle-point equation is given by $k=i\lambda$; 
$\lambda$ satisfies a self-consistency equation analogous to Eq. (\ref{tauSC}) given by
\begin{equation}\label{lambdaSC}
 \frac{I_{1}(2\lambda)}{I_{0}(2\lambda)}=\frac{\varepsilon}{2 n_{c}(\varepsilon)}.
\end{equation}
Eq.\ (\ref{ansatzUp1omega1}) can then be written as
\begin{equation}
\omega^{(2)}_{N}(\varepsilon)\simeq \omega^{(1)}_{N}(\varepsilon)e^{N\left[-\lambda\varepsilon+\log2\pi+n_c(\varepsilon)
\log\left(I_{0}(2\lambda)\right)\right]} = e^{N\left[s^{(1)}_{N}(\varepsilon)-\lambda\varepsilon+\log2\pi+n_{c}(\varepsilon)\,
 \log\left(I_{0}(2\lambda)\right)\right]}
\end{equation}
valid for $N\gg1$; $s^{(1)}_{N}(\varepsilon)$ represents is the entropy density of the two-dimensional Ising model. Dividing 
by $N$ the logarithm of the above expression, letting $N\to\infty$ and neglecting the sub-leading terms in $N$, 
we finally get the following expression for the entropy density of the $XY$ model in $d=2$
\begin{equation}\label{LMFAUp1sFINALE}
s^{(2)}(\varepsilon)\simeq s^{(1)}(\varepsilon)+\log 2\pi-\lambda\varepsilon+n_{c}(\varepsilon)\log\left[I_{0}(2\lambda)\right]\,,
\end{equation}
where $\lambda$ satisfies the self-consistency equation (\ref{lambdaSC}). 
In Fig.\ \ref{LMFAUp1energy} we plot the temperature $T$ as a function of the energy density 
$\varepsilon$ as obtained from numerical differentiation of Eq.\ (\ref{LMFAUp1sFINALE}) (red points). 
As in Fig. \ref{binomialEnergy}, the values obtained with a Monte Carlo simulation 
are shown for comparison (blue squares). 
\begin{figure}
\begin{center}
\includegraphics[width=10cm]{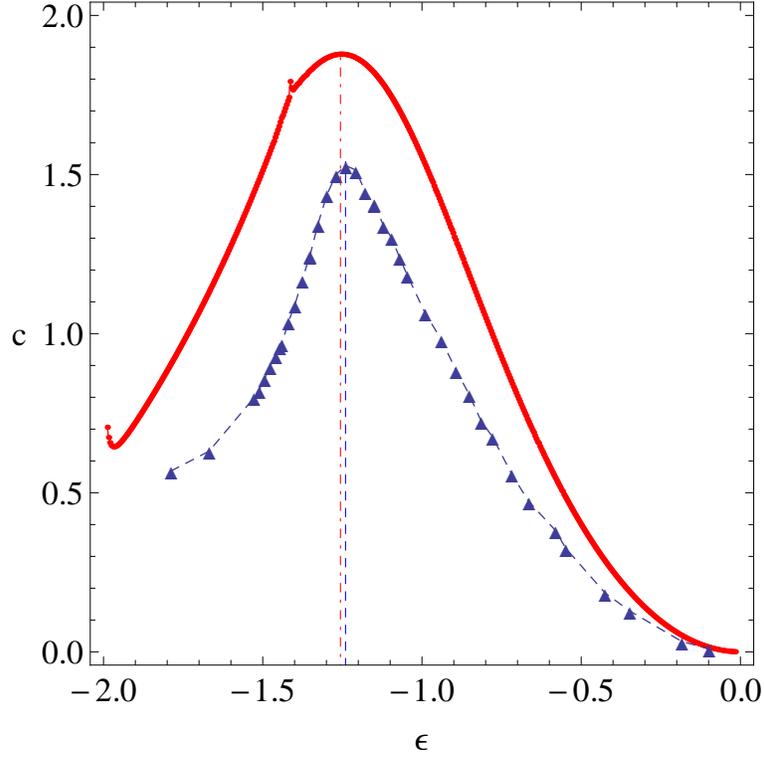}
\caption{Specific heat $c$ as a function of the energy 
density $\varepsilon$ as derived from Eq.\ (\ref{LMFAUp1sFINALE}) for the $XY$ model in $d=2$. 
Our results (red circles) are plotted together with the data obtained by a Monte Carlo simulation (blue triangles). Errorbars of the simulation data are smaller than the symbol sizes. The dashed blue line and the solid red line connecting the two sets of data are meant as guide to the eyes.}
\label{LMFAUp1Cv}
\end{center}
\end{figure}

Fig.\ \ref{LMFAUp1energy} shows that the asymptotic behavior of the function $T(\varepsilon)$ is well reproduced at a semiquantitative level by our approximation in the harmonic regime (very low-energy), in the low-energy regime and in the high-energy limit; 
the agreement is extremely good for low energies. 
For $\varepsilon\gtrsim-1.9$ the approximate results move away from the numerical ones and the approximate value of the temperature remains lower 
than the results obtained from the simulation\footnote{This behavior is analogous to the one 
obtained in \cite{Rachele:thesis} with a similar procedure applied to the partition function of the system.}. 
The largest difference between theoretical and numerical values of $T$ is about $50\%$.

In Fig.\ \ref{LMFAUp1Cv} the values of the specific heat $c$ obtained with a numerical 
differentiation of Eq.\ (\ref{LMFAUp1sFINALE}) are plotted as a function of the energy density $\varepsilon$. 
As in Fig.\ \ref{LMFAUp1energy} the theoretical results are displayed as red circles and are plotted together with 
the values of $c$ computed by Monte Carlo simulation (blue squares). 
The specific heat shows a peak for $\varepsilon_{p,2} \simeq  -1.258$ marked by the vertical red dot-dashed line. 
This value of the energy density is very close to $\varepsilon_{p} \simeq -1.24$ 
that is the energy density value at which the peak occurs in the simulation data.
Our approximation is able to reproduce the correct behavior of the specific heat in the high-energy regime, 
while the agreement becomes slightly worse in the low-energy case, although qualitatively correct. The difference 
between calculated and simulated values of $c$ is about $20\%$ for $\varepsilon<\varepsilon_{p}$ and smaller 
for $\varepsilon>\varepsilon_{p}$. The trend of the theoretical results
up to $\varepsilon\simeq-1.9$ seems to suggests a value for $c(-2)\in[0.5,0.6]$, a bit higher 
than expected. For $\varepsilon\simeq-2$ an abrupt increase of our 
results is observed as in Fig.\ \ref{BinomialCv}. Again, this is due to a shortcoming of the approximation.

The calculations presented in this Section have been repeated using the expression of 
$U_{p(\vartheta_{i})}$ given in Eq.\ (\ref{UpChoice2}). The results are presented below.

\subsection{LMF approximation for $g^{(2)}(\varepsilon)$ and $U_{p(\vartheta_{i})}$ given by Eq.\ (\ref{UpChoice2}).}
\label{subsesctionLMFAUp2}
We now consider the LMF approximation with the neighborhoods $U_{p(\vartheta_{i})}$ given by Eq.\ (\ref{UpChoice2}). 
Given an Ising configuration $p$, from Eq.\ (\ref{LMFAfact1}) we have two possibilities.

$(i)$ if $\bar{\vartheta}^{(1)}_{i}=\bar{\vartheta}^{(2)}_{i}=1$ (resp.\ $-1$) and 
$\bar{\vartheta_{i}}=1$ (resp. $-1$), i.e., the 
configuration locally looks like
\begin{equation}\label{e:local_config1}
\begin{matrix}
\cdot & \uparrow & \cdot & \qquad & \qquad & \qquad & \cdot & \downarrow & \cdot \\
\cdot & \uparrow & \uparrow & \qquad & \mbox{or respectively} & \qquad & \cdot &\downarrow &\downarrow\;,\\
\cdot & \cdot & \cdot & \qquad & \qquad & \qquad & \cdot &\cdot &\cdot
\end{matrix}
\end{equation} 
then
\begin{equation}\label{LMFAUp2a}
\int_{-\frac{\pi}{2}}^{\frac{\pi}{2}}e^{-ik\cos\vartheta_{i} (\cos\bar{\vartheta}^{(1)}_{i}+
\cos\bar{\vartheta}^{(2)}_{i})} d\vartheta_{i} = 
\int_{\frac{\pi}{2}}^{\frac{3\pi}{2}}e^{-ik\cos\vartheta_{i} (\cos\bar{\vartheta}^{(1)}_{i}+
\cos\bar{\vartheta}^{(2)}_{i})} d\vartheta_{i}= \pi\left(J_{0}(2k)-iH_{0}(2k)\right)
\end{equation}
regardless of whether $\bar{\vartheta}_{i}=0$ or $\pi$; $H_{0}(x)$ denotes the zero-order Struve function.

$(ii)$ If $\bar{\vartheta}^{(1)}_{i}=\bar{\vartheta}^{(2)}_{i}=1$ (resp. $-1$) and 
$\bar{\vartheta_{i}}=-1$ (resp. $1$), i.e., the 
configuration locally looks like
 \begin{equation}\label{e:local_config2}
\begin{matrix}
\cdot & \uparrow & \cdot & \qquad & \qquad & \qquad & \cdot & \downarrow & \cdot \\
\cdot & \downarrow & \uparrow & \qquad & \mbox{or respectively} & \qquad & \cdot &\uparrow &\downarrow\;,\\
\cdot & \cdot & \cdot & \qquad & \qquad & \qquad & \cdot &\cdot &\cdot
\end{matrix}
\end{equation}
then
\begin{equation}\label{LMFAUp2b}
\int_{-\frac{\pi}{2}}^{\frac{\pi}{2}}e^{-ik\cos\vartheta_{i} (\cos\bar{\vartheta}^{(1)}_{i}+
\cos\bar{\vartheta}^{(2)}_{i})} d\vartheta_{i} =
\int_{\frac{\pi}{2}}^{\frac{3\pi}{2}}e^{-ik\cos\vartheta_{i} (\cos\bar{\vartheta}^{(1)}_{i}+
\cos\bar{\vartheta}^{(2)}_{i})} d\vartheta_{i}= \pi\left(J_{0}(2k)+iH_{0}(2k)\right)
\end{equation}
regardless of whether $\bar{\vartheta}_{i}=0$ or $\pi$.

On the other hand, Eq.\ (\ref{LMFAfact2}) is simply given by 
\begin{equation}
\int_{-\frac{\pi}{2}}^{\frac{\pi}{2}}d\vartheta_{i}\;1=\int_{\frac{\pi}{2}}^{\frac{3\pi}{2}}d\vartheta_{i}\;1=\pi\,.
\end{equation}
We will denote by $n_{3}=N_{3}/N$ the density of triplets of equal spins forming a local configuration as in Eq.\ (\ref{e:local_config1}). Combining Eqs.\ (\ref{LMFAUp2a}) and (\ref{LMFAUp2b}) with Eq.\ (\ref{LMFA}) we get
\begin{equation}\label{G2Up2step1}
g^{(2)}(\varepsilon)=\frac{\pi^{N}}{2\pi}\int_{-\infty}^{\infty}dk\;e^{ikN\varepsilon}\; e^{N\left[n_{3}(\varepsilon)
\log\left(J_{0}(2k)-iH_{0}(2k)\right)\right]}
\qquad\qquad\times  e^{N\left[(n_{c}(\varepsilon)-n_{3}(\varepsilon))\log\left(J_{0}(2k)+iH_{0}(2k)\right) \right]}.
\end{equation}
In this case, the vector of parameters $\mathbf{z}(2,2,\varepsilon)$ is given by 
$\mathbf{z}(2,2,\varepsilon)=n_{c}(\varepsilon)\cup n_{3}(\varepsilon)$.

The integral in Eq.\ (\ref{G2Up2step1}) can be computed with the saddle point method in the large-$N$ limit. 
The saddle point equation is given by $k=-i\zeta$; 
making use of the properties of the Bessel and of the Struve functions and performing some algebra,
Eq.\ (\ref{G2Up2step1}) can be written as
\begin{equation}\label{G2Up2step2}
g^{(2)}(\varepsilon)\simeq e^{N\left[-\zeta\varepsilon + \log\pi + n_{3}(\varepsilon) 
\log\left(I_{0}(2\zeta) - L_{0}(2\zeta) \right)\right]} e^{N\left[\left(n_{c}(\varepsilon)-n_{3}(\varepsilon) \right) \log\left(I_{0}(2k)+L_{0}(2\zeta) \right) \right]}
\end{equation}
with $\zeta$ satisfying the self-consistency equation
\begin{equation}\label{zetaSC}
\begin{split}
-&\varepsilon + \frac{n_{3}(\varepsilon)}{I_{0}(2\zeta)-L_{0}(2\zeta)}\left( 2I_{1}(2\zeta)-
\left( \frac{2}{\pi}+L_{-1}(2\zeta)+L_{1}(2\zeta) \right) \right)+\\
+&\frac{n_{c}(\varepsilon)-n_{3}(\varepsilon)}{I_{0}(2\zeta)+L_{0}(2\zeta)}\left( 2I_{1}(2\zeta)+
\left( \frac{2}{\pi}+L_{-1}(2\zeta)+L_{1}(2\zeta) \right) \right)=0.
\end{split}
\end{equation}
As in the case of $n_{c}(\varepsilon)$, the function $n_{3}(\varepsilon)$ can be obtained  interpolating the numerical data arising from a Monte Carlo simulation of the system; the simulation data for $n_{3}$ are not shown here.

We can now insert Eq.\ (\ref{G2Up2step2}) in Eq.\ (\ref{sec1:omega_appr_richiamo}). 
By taking the logarithm of the resulting expression and neglecting the sub-leading term in $N$, 
we finally arrive to the following expression for the entropy density
\begin{equation}\label{LMFAUp2sFINALE}
s^{(2)}(\varepsilon)\simeq \,s^{(1)}(\varepsilon)\, -\, \zeta\, \varepsilon +\, \log \pi \,+
\,n_{3}(\varepsilon)\,\log\left[I_{0}(2\zeta)-L_{0}(2\zeta)\right]\,+
(n_{c}(\varepsilon)-n_{3}(\varepsilon))\,\log\left[I_{0}(2\zeta)+L_{0}(2\zeta)\right]
\end{equation}
valid in the $N\to\infty$ limit; $\zeta$ has to be determined numerically from Eq.\ (\ref{zetaSC}).
\begin{figure}
\centering
\includegraphics[width=10cm]{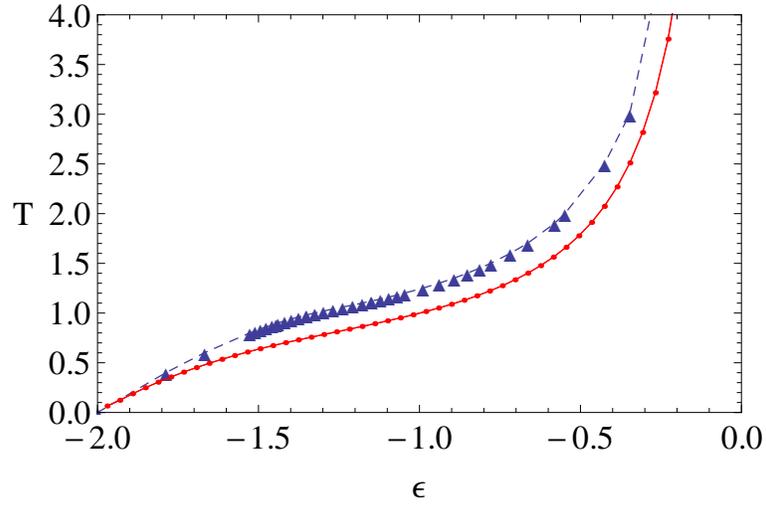} 
\caption{Temperature $T$ as a function of the energy density $\varepsilon$ as derived from 
Eq.\ (\ref{LMFAUp2sFINALE}) for the $XY$ model in $d=2$. 
Our results (red circles) are plotted together with the data obtained by a Monte Carlo simulation (blue triangles). Errorbars of the simulation data are smaller than the symbol sizes. 
The dashed blue line and the solid red line connecting the two sets of data are meant as guide to the eyes.}
\label{LMFAUp2energy}
\end{figure}

\begin{figure}
\begin{center}
\includegraphics[width=10cm]{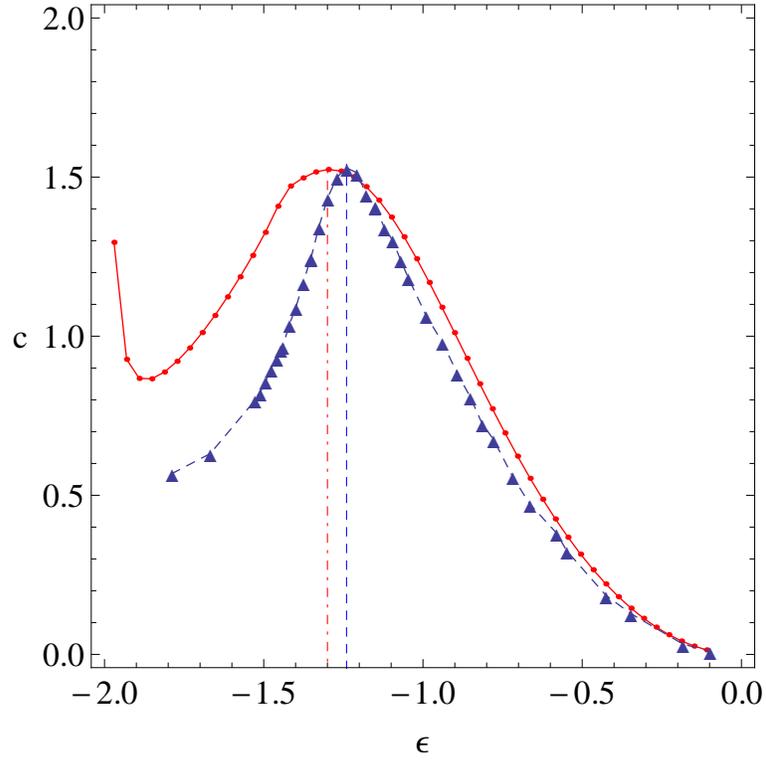}
\caption{Specific heat $c$ as a function of the energy density $\varepsilon$ as 
derived from Eq. (\ref{LMFAUp1sFINALE}) for the $XY$ model in $d=2$. Our results (red circles) are plotted 
together with the data obtained by a Monte Carlo simulation (blue triangles). Errorbars of the simulation data are smaller than the symbol sizes. 
The dashed blue line and the solid red line connecting the two sets of data are meant as guide to the eyes.}
\label{LMFAUp2Cv}
\end{center}
\end{figure}
Fig.\ \ref{LMFAUp2energy} shows $T$ as function of the energy density $\varepsilon$ obtained by numerical differentiation from Eq.\ (\ref{LMFAUp2sFINALE}) (red points). As in Fig.\ \ref{LMFAUp1energy} the theoretical 
values are plotted together with the data obtained by a Monte Carlo simulation of the system (blue squares). 

For Fig.\ \ref{LMFAUp2energy} the same comments can be done as for Fig.\ \ref{LMFAUp1energy}. 
The theoretical results are in good agreement with the numerics in the entire energy density range. 
In comparison with the theoretical caloric curve in Fig.\ \ref{LMFAUp1energy}, the caloric curve resulting from 
this approximation and shown in Fig. \ref{LMFAUp2energy} is closer to the numerical results: the largest difference between 
theory and simulation is here around $20\%$. This fact is the effect 
of the different choice of the integration neighborhoods. In particular, if the integration neighborhoods 
are superposed, as in Eq.\ (\ref{UpChoice1}), $g^{(2)}(\varepsilon)$ is overestimated 
and the difference between the theoretical and the numerical results is 
larger than in the case of a choice of the integration neighborhoods as in Eq.\ (\ref{UpChoice2}).

This fact is even more evident for the specific heat. The theoretical results are plotted in red in Fig.\ \ref{LMFAUp2Cv}. 
With a choice of the integration neighborhoods of Eq. (\ref{G2Up2step1}) as in Eq.\ (\ref{UpChoice2}),
the energy value of the peak of the specific heat derived with our approximation is 
$\varepsilon_{p,3} \simeq -1.3\simeq \varepsilon_{p} \simeq -1.24$;  
moreover, the entire high energy regime for $\varepsilon>\varepsilon_{p}$ is in good quantitative agreement with the 
numerics. On the other hand, for $\varepsilon<\varepsilon_{p}$ the two sets of data separate themselves and in the low energy 
regime the same considerations as for Figs.\ \ref{BinomialCv} and \ref{LMFAUp1Cv} can be done. 

\subsection{Harmonic approximation for $g^{(2)}(\varepsilon)$}\label{sub:harmonic}
In this last Section we present an alternative way to estimate the function $g^{(2)}$ appearing 
in Eq.\ (\ref{sec1:omega_appr_richiamo}), using a harmonic expansion of the Hamiltonian around each Ising stationary point. 
This approach can be seen as the natural generalization of the classical low temperature harmonic expansion, see e.g.\ the Stillinger and Weber theory \cite{StillingerWeber:science1984}. However, in the present case, 
Ising stationary points are not only minima but saddles of every index so that the generalization is far from 
being straightforward. For this reason, we only discuss the ``ansatz-based'' approach here.

Let us consider for the moment only the factor $g^{(2)}(\varepsilon, \bar{p})$ 
in Eq.\ (\ref{continuous-factor}). The approximation consists in expanding the Hamiltonian up to the harmonic order 
around the Ising configuration $\bar{p}=(\bar{\vartheta}_1,...\bar{\vartheta}_N)$:
\begin{equation}
g^{(2)}(\varepsilon, \bar{p})\simeq g^{(2)}_{har}(\varepsilon, \bar{p})=
\int_{U_{\bar{p}}}d\vartheta_{1}...d\vartheta_{N}\,\delta\left(\frac{1}{2}(\vartheta - \bar{p})\,
\mathcal{H}(\bar{p})\,(\vartheta - \bar{p})^{T}\,-\, N(\varepsilon-\varepsilon_{\bar{p}})\right)\,,
\end{equation}
where $\mathcal{H}(\bar{p})$ is the Hessian matrix of $H^{(2)}$ evaluated on the Ising stationary point $\bar{p}$ 
with energy $H^{(2)}(\bar{p})=N\varepsilon_{\bar{p}}$. We now shift the coordinates to move the stationary 
point $\bar{p}$ to the origin of the coordinate system, a change of variables to diagonalize the Hessian matrix and 
a rescaling of the integration variables according to the eigenvalues of the Hessian. With this procedure, we have
\begin{equation}
g^{(2)}(\varepsilon, \bar{p})\simeq g^{(2)}_{har}(\varepsilon, \bar{p})=\frac{1}{\sqrt{2^{N}\mathcal{D}(\bar{p})}}
\int_{U_{\bar{p}}}dy_{_{1}}...dy_{_{N_{+}}}\, dx_{_{1}}...dx_{_{N_{-}}}\,
\delta\left(\sum_{i=1}^{N_{+}(\bar{p})}y_{i}^{2}-\sum_{i=1}^{N_{-}(\bar{p})}x_{i}^{2}-\, 
N(\varepsilon-\varepsilon_{\bar{p}})\right)\label{eq:1}
\end{equation}
where $\mathcal{D}(\bar{p})$ is the determinant of $\mathcal{H}(\bar{p})$, $N_-(\bar{p})$ is the index of $\bar{p}$, 
$N_+(\bar{p})=N-N_-(\bar{p})$. Observe that in the previous expression, $y_i$ with $1\leq i\leq N_+(\bar{p})$ are 
the variables associated with the directions in phase space corresponding to the positive eigenvalues and $x_i$ with $1\leq i\leq N_-(\bar{p})$ are the variables associated with the directions in the phase space 
corresponding to the negative eigenvalues. 

To go further, we have to specify in a suitable way the integration neighborhoods $U_{\bar{p}}$. 
As in the LMF case, this choice determines the feasibility of the calculation. In the following, 
we consider $U_{\bar{p}}$ as the union of two balls \footnote{A different choice of the 
$U_{\bar{p}}$ given by a single ball, may look more natural at first sight. This choice, however, 
lead to the inconsistent result that if $N_-=0$, the minimum energy density value available to the system in the set 
$U_{\bar{p}}$ is lower than $\varepsilon_{\bar{p}}$.} centered on the stationary point. The first one is associated 
with the directions $y_i$ and has radius $\alpha$, and the second one to the directions $x_i$ and has radius $\beta$. 
With this choice, we have
\begin{equation}
g^{(2)}_{har}(\varepsilon, \bar{p})=\frac{1}{\sqrt{2^{N}\mathcal{D}(\bar{p})}}
\int_{-\infty}^{\infty}dy_{_{1}}...dy_{_{N_{+}}}\, dx_{_{1}}...dx_{_{N_{-}}}\,\delta\left(Y^2-X^2-\, 
N(\varepsilon-\varepsilon_{\bar{p}})\right)\Theta\left(-Y^{2}+N_{+}\alpha\right)\Theta\left(-X^{2}+N_{-}\beta\right)\,.
\label{eq:2ball-integration-1}
\end{equation}
where we have introduced the following variables:
\begin{equation}
Y^{2}=\sum_{i=1}^{N_{+}(p)}y_{i}^{2}\qquad\textrm{and}\qquad X^{2}=\sum_{i=1}^{N_{-}(p)}x_{i}^{2}\,.\label{eq:X-Y-variables}
\end{equation}
In principle,  $\alpha$ and $\beta$ should be considered as functions of the specific Ising configuration 
$\bar{p}$. This would however introduce in the theory a huge number of free parameters. 
Thus, we will assume that $\alpha$ and $\beta$ are fixed real numbers. 
Moreover, $g^{(2)}_{har}$ depends only very weakly on $\alpha$: this can be easily understood by recalling that, 
in the canonical ensemble, large values in the positive directions are damped in a Gaussian way. 
For this reason, we can also assume that $\alpha\gg \beta$.

As explained in Section \ref{subsec:degeneracyfactor}, in the ``ansatz-based'' approach the continuous weight does not 
really depend on the specific Ising stationary point $\bar{p}$ but only on its energy density, 
which corresponds to assume that  
\begin{equation}
\left\{\mathcal{D}(\bar{p})^{\frac{1}{N}},\frac{N_+(\bar{p})}{N},\frac{N_-(\bar{p})}{N}\right\}\simeq
\left\{\mathcal{D}(\varepsilon)^{\frac{1}{N}},\frac{N_+(\varepsilon)}{N},\frac{N_-(\bar{\varepsilon})}{N}\right\}\,. 
\label{eq:harm-similarity}
\end{equation}
These functions can be computed numerically as in the case of $n_{c}(\varepsilon)$ and $n_{3}(\varepsilon)$ discussed in 
Secs.\ \ref{FPA} and \ref{ansatz}, respectively (data not shown). 
We recall that in the ``ansatz-based'' approach, only continuous weights corresponding to Ising stationary points with energy $\varepsilon_{\bar{p}}=\varepsilon$ contribute to $\omega^{(2)}$. 
We refer the reader to Appendix \ref{app:harmonic} for the detailed computation of $g^{(2)}$. 
Under these conditions, the approximate density of states becomes
\begin{equation}
\omega^{(2)}(\varepsilon)\simeq\exp^{\left\{ \frac{N}{2}\left[2s^{(1)}(\varepsilon)+ 1-\log2-
\mathcal{LD}(\varepsilon)+\log\pi-n_{+}(\varepsilon)\log\frac{n_{+}(\varepsilon)}{2}-
n_-(\varepsilon)\log\frac{n_{-}(\varepsilon)}{2}+\log\left(n_{-}(\varepsilon)\right)+\log\left(\beta\right)\right]\right\}} 
\label{eq:harmonic-final}
\end{equation}
where $\mathcal{LD}=\lim_{N\to\infty}\frac{1}{N}\log \mathcal{D}$, $n_-=N_-/N$, $n_+=1-n_-$. 
Two observations are mandatory at this level. First, we note that 
Eq.\ (\ref{eq:harmonic-final}) does not depend anymore on $\alpha$; second, even if the entropy does depend on $\beta$, 
thermodynamic functions such as temperature or specific heat do not depend on the free parameter $\beta$. Thus, 
thermodynamic functions do not depend on any free parameters.

\begin{figure}
\centering
\includegraphics[width=10cm]{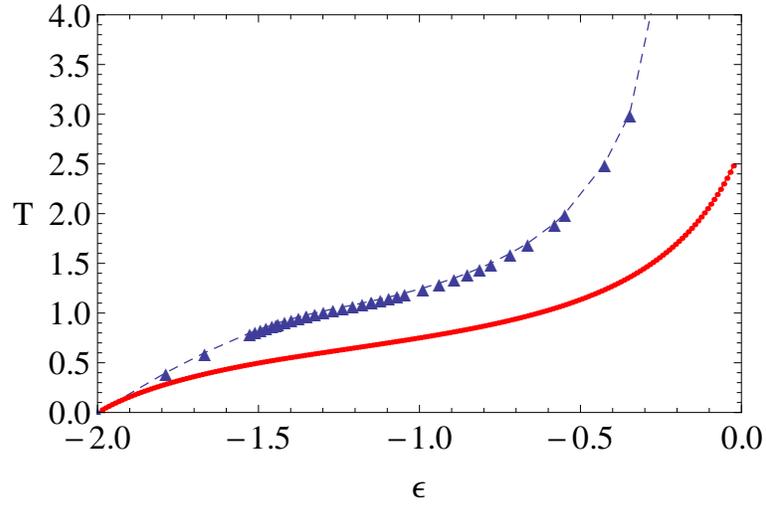} 
\caption{Temperature $T$ as a function of the energy density $\varepsilon$ as derived 
from Eq.\ (\ref{eq:harmonic-final}) for the $XY$ model in $d=2$. 
Our results (red circles) are plotted together with the data obtained by a Monte Carlo simulation (blue triangles). Errorbars of the simulation data are smaller than the symbol sizes. 
The dashed blue line and the solid red line connecting the two sets of data are meant as guide to the eyes.}
\label{harmonic-energy}
\end{figure}

\begin{figure}
\begin{center}
\includegraphics[width=10cm]{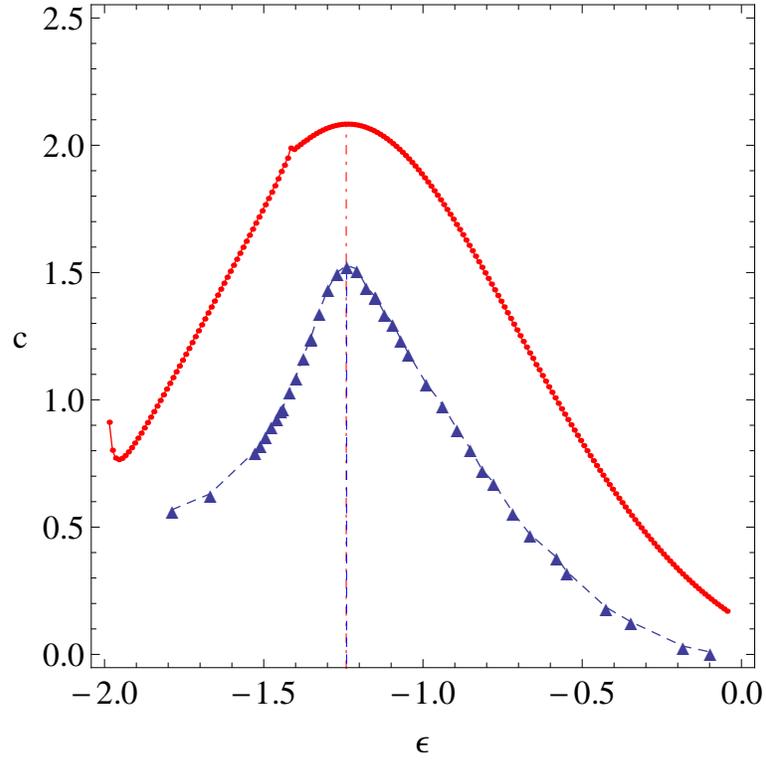}
\caption{Specific heat $c$ as a function of the energy 
density $\varepsilon$ as derived from Eq.\ (\ref{eq:harmonic-final}) for the $XY$ model in $d=2$. 
Our results (red circles) are plotted together with the data obtained by a Monte Carlo simulation (blue triangles). Errorbars of the simulation data are smaller than the symbol sizes. 
The dashed blue line and the solid red line connecting the two sets of data are meant as guide to the eyes.}
\label{harmonic-cv}
\end{center}
\end{figure}

The behavior of the temperature and of the specific heat as a function of the energy density $\varepsilon$ derived from 
Eq.\ (\ref{eq:harmonic-final}) are reported in Fig.\ \ref{harmonic-energy} and \ref{harmonic-cv}, respectively. For the latter figures, 
similar considerations as in the case of Fig.\ \ref{LMFAUp1energy} and \ref{LMFAUp1Cv} can be done. 

\section{Concluding remarks}\label{ch5fine}
In this paper, we have presented two different semi-analytic ways to approximate the density of states, hence thermodynamic functions, of short-range O$(n)$ models on a regular lattice. We have performed explicit calculations in the case of the $XY$ model in $d=2$, but the generalization of our results to higher dimensions and/or to O$(n)$ models with $n>2$ should be possible along the same lines. We used an energy landscape approach inspired by our previous works reported in Refs.\ \cite{prl2011,jstat2012}. 

We have developed two different schemes of approximation. The first one, called ``first-principle'' approximation, is an attempt to follow the same kind of derivation used for the $1$-$d$ and the mean-field $XY$ models in \cite{jstat2012}. The second one, called ``ansatz-based'' approximation, assumes Eq.\ (\ref{intro:ansatz}) as an ansatz on the form of the density of states of O$(n)$ models with $n\geq2$.
In both cases, the delicate point is to approximate the weights $g$ given by Eq.\ (\ref{continuous-factor}). Two different approximations have been used, the local mean field (LMF) and a harmonic approximation around each Ising configuration. 

The $XY$ model in $d=2$ is not exactly solvable and any approximation scheme has to be compared with the results coming from numerical simulations. 
Calculated thermodynamic functions always show a qualitative agreement with the numerical data and in some cases also a quantitative agreement. 
Particularly interesting is the presence of the peak in the specific heat reported in Figs.\  \ref{BinomialCv}, \ref{LMFAUp1Cv} 
and \ref{LMFAUp2Cv}. Despite the approximations involved in its derivation, the specific heat correctly shows a peak and not a divergence as it happens, instead, for the Ising model in $d=2$ at $\varepsilon^{(1)}_{c} = \sqrt{2}$. 
For this particular value of the energy density our numerical procedure correctly produces a finite value of $c$ although the
numerical convergence is more delicate as highlighted by the scattered data present in Figs.\  \ref{LMFAUp1Cv}  and \ref{LMFAUp2Cv} for $\varepsilon\simeq\varepsilon^{(1)}_{c}$. In case of Fig.\ \ref{LMFAUp2Cv} the agreement is also 
quantitative as far as the location of the peak of the specific heat is concerned. 

As stressed throughout the paper, the concepts presented here are valid in principle for 
any O$(n)$ model in any spatial dimensions. Hence the generalization of the calculations carried out here for the $XY$ model on a square lattice to other O$(n)$ models in $d$ dimensions should be possible, with possibly some technical complications, and may give a hint towards the development of approximation techniques that may be valid for estimating the density of states of large classes of Hamiltonian systems.


\appendix
\section{``Ansatz-based'' approximation with harmonic expansion}
\label{app:harmonic}
In this Appendix, we compute exactly the integrals in Eq.\ (\ref{eq:2ball-integration-1}) under the assumption $\varepsilon_{\bar{p}}=\varepsilon$. We observe that the computation in this appendix can be extended also to the case $\varepsilon_p\neq \varepsilon$; however, we do not report the details of this more general case here, as we do not need it for the development of the ``ansatz-based'' approximation.

Inserting the change of measure in Eq.\ (\ref{eq:2ball-integration-1}) due to the change of variables 
in Eq. (\ref{eq:X-Y-variables}), and setting $\varepsilon_p= \varepsilon$, we have
\begin{equation}
g^{(2)}_{har}(\varepsilon, \bar{p})=\frac{C(N_+)C(N_-)}{\sqrt{2^{N}\mathcal{D}(\bar{p})}}
\int_{0}^{\infty}dY\,dX\,Y^{N_+-1}\,X^{N_--1}\,\delta\left(Y^2-X^2\right)\Theta\left(-Y^{2}+N_{+}\alpha\right)
\Theta\left(-X^{2}+N_{-}\beta\right)\,,\label{eq:2ball-integration-2}
\end{equation}
where $C(k)$ is the surface of the unitary $(k-1)$-dimensional sphere
$\mathbb{S}^{k-1}$:
\begin{equation}
C(k)=\frac{k\pi^{\frac{k}{2}}}{\Gamma\left(\frac{k}{2}+1\right)}\,.
\end{equation}
Using the delta functional to express $X$ as a function of $Y$ and with simple computations, we get
\begin{eqnarray}
g^{(2)}_{har}(\varepsilon, \bar{p})&=&\frac{C(N_+)C(N_-)}{\sqrt{2^{N}\mathcal{D}(\bar{p})}}\int_{0}^{\infty}dY\,Y^{N-2}\,
\Theta\left(-Y^{2}+N_{+}\alpha\right)\Theta\left(-Y^{2}+N_{-}\beta\right)\,=\nonumber\\
&=&\frac{C(N_+)C(N_-)}{\sqrt{2^{N}\mathcal{D}(\bar{p})}}\int_{0}^{\sqrt{N\,n_-\beta}}dY\,Y^{N-2}=\\
&=&\frac{1}{N-1}\frac{C(N_+)C(N_-)}{\sqrt{2^{N}\mathcal{D}(\bar{p})}}(N\,n_-\beta)^{\frac{N-1}{2}}\,.
\label{eq:2ball-integration-3}\nonumber
\end{eqnarray}
To pass from the first to the second line we assumed that $\alpha\gg\beta$, as discussed in Section 
\ref{sub:harmonic}, so that $\textrm{min}\{n_+\alpha,n_-\beta\}= n_-\beta$.
The above expression for $g^{(2)}_{har}$ can be simplified in the large $N$ limit by retaining only those 
factors that are exponential in $N$. We thus obtain
\begin{eqnarray}
g^{(2)}_{har}(\varepsilon, \bar{p})\sim_{N\gg1}\exp\left\{ \frac{N}{2}\left[ 1-\mathcal{LD}(\bar{p})+
\log\frac{\pi}{2}-n_{+}(\bar{p})\log\frac{n_{+}(\bar{p})}{2}-n_-(\bar{p})\log\frac{n_{-}(\bar{p})}{2}+
\log\left(n_{-}(\bar{p})\right)+\log\left(\beta\right)\right]\right\}\,,
\end{eqnarray}
where we have used the classical asymptotic expansion of the Gamma function. Inserting the last expression 
in the ansatz (\ref{sec1:omega_appr_richiamo}), making use of Eq.\ (\ref{eq:harm-similarity}), we obtain the approximate 
density of states given in Eq.\ (\ref{eq:harmonic-final}).

\bibliography{/Users/casetti/Work/Scripta/papers/bib/mybiblio,/Users/casetti/Work/Scripta/papers/bib/statmech}

\end{document}